\makeatletter 

\documentclass[aps,twocolumn,pra,tightenlines,floatfix,showpacs,superscriptaddress]{revtex4-1}

\usepackage{graphicx}
\usepackage{epstopdf}
\usepackage[english]{babel}
\usepackage{amsmath}
\usepackage{ntheorem}
\theoremseparator{:}
\theoremheaderfont {\it}

\usepackage{amssymb}
\usepackage{times}
\usepackage{appendix}
\usepackage{bm}
\usepackage{mathrsfs}   
\usepackage{float}
\usepackage{color}
\usepackage{longtable}
\usepackage{url}
\usepackage[usenames,dvipsnames]{xcolor}
\usepackage{subfigure}
\usepackage[percent]{overpic}
\renewcommand{\@thesubfigure}{\hskip\subfiglabelskip}
\usepackage{indentfirst} 
\graphicspath{{./Figures}}

\newcommand{\Rmnum}[1]{\expandafter\@slowromancap\romannumeral #1@}

\usepackage{array}
\usepackage{booktabs}

\newcommand{\rmi}{\text{i}}
\newcommand{\ket}[1]{\left|#1\right\rangle}

\newcommand{\n}{\nonumber\\}

\newcommand{\ex}[1]{\langle #1\rangle}

\makeatother 

\usepackage[colorlinks=true,linkcolor=Blue,urlcolor=Blue,citecolor=Blue,anchorcolor=Red]{hyperref}
\usepackage{cleveref}
\crefname{equation}{Eq.~}{Eq.~}
\crefname{figure}{Fig.~}{Fig.~}

\begin{document}

\title{Quantum metrology enhanced by the $XY$ spin interaction in a generalized Tavis-Cummings model}

%

\author{Yuguo Su}

\email{suyuguo@apm.ac.cn}

\affiliation{Innovation Academy for Precision Measurement Science and Technology, Chinese Academy of Sciences, Wuhan 430071, China}

\author{Wangjun Lu}

\affiliation{Department of Maths and Physics, Hunan Institute of Engineering, Xiangtan 411104, China}

\author{Hai-Long Shi}

\email{hl\_shi@yeah.net}

\affiliation{QSTAR and INO-CNR, Largo Enrico Fermi 2, 50125 Firenze, Italy}
\affiliation{Hefei National Laboratory, Hefei 230088 China}

\begin{abstract}
Quantum metrology is recognized for its capability to offer high-precision estimation by utilizing quantum resources, such as quantum entanglement.
Here, we propose a generalized Tavis-Cummings model by introducing the $XY$ spin interaction to explore the impact of the many-body effect on estimation precision, quantified by the quantum Fisher information (QFI).
By deriving the effective description of our model, we establish a closed relationship between the QFI and the spin fluctuation induced by the $XY$ spin interaction.
Based on this exact relation, we emphasize the indispensable role of the spin anisotropy in achieving the Heisenberg-scaling precision for estimating a weak magnetic field.
Furthermore, we observe that the estimation precision can be enhanced by increasing the strength of the spin anisotropy. 
We also reveal a clear scaling transition of the QFI in the Tavis-Cummings model with the reduced Ising interaction.
Our results contribute to the enrichment of metrology theory by considering many-body effects, and they also present an alternative approach to improving the estimation precision by harnessing the power provided by many-body quantum phases.

\end{abstract}
\maketitle


\section{Introduction}\label{Sec.I}
Quantum metrology~\cite{PhysRevLett.96.010401,WOS:000288984900012,RevModPhys.89.035002,RevModPhys.90.035005,Carollo_2019,10.21468/SciPostPhys.13.4.077,PhysRevLett.128.160505} aims to achieve enhanced sensitivity in estimating an unknown parameter $\xi$ compared to classical metrology by exploiting diverse quantum resources, such as quantum entanglement~\cite{PhysRevA.54.R4649,PhysRevLett.102.100401,PhysRevLett.106.130506,PhysRevLett.107.080504,PhysRevA.85.022321}  and quantum squeezing~\cite{PhysRevD.23.1693,PhysRevLett.88.231102,MA201189,PhysRevLett.110.163604,PhysRevA.92.023603,PhysRevA.102.052423}.
The widespread applications of quantum metrology have emerged in numerous experimental domains, including Ramsey spectroscopy~\cite{WOS:A1980KA25400008,PhysRevLett.86.5870}, atomic clocks~\cite{RevModPhys.87.637,Louchet_Chauvet_2010,PhysRevLett.112.190403}, gravitational-wave detectors~\cite{WALLS1981118,WOS:000256613000015,Abbott_2009}, magnetometry~\cite{budker_optical_2007,PhysRevLett.109.253605,PhysRevLett.120.260503,PhysRevLett.126.010502}, and biophysical  measurements~\cite{doi:10.1073/pnas.1004037107,WOS:000316154700018,TAYLOR20161}.
The quantum Cram\'er-Rao bound (QCRB)~\cite{PhysRevLett.72.3439} $\delta \xi\geq1/\sqrt{\mathcal{F}_{\xi}}$ provides a theoretical approach to assess the suitability of a quantum state for a high-precision quantum estimation.
In other words, a quantum state with a larger quantum Fisher information (QFI), $\mathcal F_\xi$, can yield a higher precision.
For example, in the quantum phase estimation, when utilizing non-entangled $N$-particle states, the precision is bounded by the standard quantum limit (SQL), namely, $\mathcal F_{\xi}\propto N$.
However, when the Greenberger-Horne-Zeilinger (GHZ) entangled state is employed, it becomes feasible to approach the Heisenberg limit (HL), meaning that $\mathcal F_{\xi}\propto N^2$.
Preparing a specified GHZ state is, however, a challenging endeavor from an experimental perspective.
It is also unclear whether it has the capability to sense a non-global parameter, such as the strength of a magnetic field, with high precision.
Through the adiabatic state preparation technique, it becomes possible to prepare more many-body ground states beyond the GHZ state \cite{PhysRevLett.100.030502,PhysRevA.81.061603,Ciavarella2023statepreparationin,PhysRevA.71.053814,PRXQuantum.3.020347,PhysRevLett.131.060602}.
Therefore, our aim is to explore whether certain many-body ground states, such as the ground state of the $XY$ spin chain, are sufficient to achieve HL-precision metrology.

Cavity quantum electrodynamics (Cavity-QED) contributes significantly to our foundational understanding of the interaction between atoms and the electromagnetic field within a cavity~\cite{Dutra-Book:2005,Raimond-RMP:2001,hepp1973superradiant,PhysRevA.8.2517}.
Its remarkable capability to manipulate atoms using the electromagnetic field has proven to be highly valuable across a wide spectrum of domains, spanning from quantum simulation~\cite{Buluta2009,Georgescu2014}, quantum key distribution~\cite{Scarani2009,obrien_photonic_2009}, generation of nonclassical correlations~\cite{PhysRevLett.80.3948,PhysRevA.92.013830,Stassi_2016}, quantum batteries~\cite{campaioli2023colloquium,PhysRevLett.120.117702,PhysRevB.105.115405,PhysRevLett.129.130602,PhysRevA.109.012204}, to quantum metrology~\cite{,WOS:000288984900012,Liu_2019,YuguoSu2021,PhysRevLett.130.170801,deng2023,WOS:000368673800032,PhysRevLett.116.093602,Flower_2019,Gietka_2021,Kim1999,WOS:000263818900002,PhysRevA.94.022313,wan2023quantum}.
The Tavis-Cummings (TC) model~\cite{PhysRev.170.379,PhysRev.188.692} is the simplest one for describing a multi-atom cavity quantum electrodynamics system.
To enhance the many-body effects within the TC model, we extend our consideration to a generalized TC model in which the atoms are organized into a spin chain with the $XY$ interaction.
Our primary focus will be on investigating variations in metrological precision when the initial state of the atoms is prepared under different quantum phases of the $XY$ spin chain.

In this work, a promising quantum-enhanced protocol is proposed for sensing a weak magnetic field in a cavity-QED system by introducing the many-body spin effect.
We establish a direct relationship between the estimation precision and the spin fluctuations induced by the $XY$ spin interaction.
Moreover, based on this exact relation, we illustrate that the spin anisotropy is pivotal in achieving the HL-precision  for estimating a weak magnetic field, and elucidate that the estimation precision can be enhanced by increasing the strength of the spin anisotropy. 
Additionally, we utilize the estimation precision to identify quantum phase transitions in the TC model with the reduced Ising interaction. 
Our work provides an effective attempt at designing high-precision quantum estimation strategies by incorporating the quantum many-body effect.


This paper is organized as follows.
In Sec.~\ref{Sec.II}, we introduce a generalized TC model featuring the $XY$ spin interaction.
The effective description of our model is obtained by using the time-averaged method and validated through numerical calculations.
In Sec.~\ref{Sec.III}, based on the effective Hamiltonian, we analytically derive the QFI and demonstrate the indispensability of the spin fluctuation for achieving the HL-precision.
By utilizing the correlation function of the $XY$ model, we analyze the scaling of the QFI within different quantum phases. 
We show that the TC model, devoid of any spin interactions, achieves only the SQL-precision.
Conversely, when incorporating the $XY$ spin interaction and the spin anisotropy into the TC model, the HL-precision becomes attainable for estimating a weak magnetic field.
We elucidate a distinct scaling transition of the QFI in the TC model with the Ising interaction.
Furthermore, we elaborate on the effect of the general $XY$ spin interaction and illustrate that the QFI increases with the raise of the spin anisotropy.
Finally, a conclusion is made in Sec.~\ref{Sec.V}.

\begin{figure}[htbp]
	\centering
	\begin{minipage}{1\linewidth}
		\centering
		\begin{overpic}[width=0.95\linewidth]{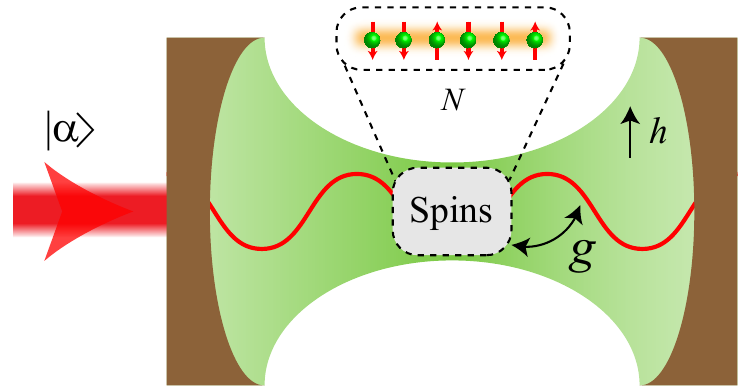}
		\end{overpic}
	\end{minipage}
	\caption{
		(Color online) 
		Schematic illustration of the generalized TC model: A coherent light $\left|\alpha\right\rangle$ is injected into the cavity and coupled to $N$ trapped spins (depicted in green) with a coupling strength $g$. 
		These $N$ spins collectively form the $XY$ spin chain and are employed for sensing a magnetic field $h$.
	}\label{Fig1}
\end{figure}

\section{System and effective Hamiltonian}\label{Sec.II}

We consider a generalized TC model, where a collection of $N$ two-level atoms (or spin-1/2 spins) constitutes a $XY$ spin chain and collectively interacts with a single bosonic cavity mode,  as illustrated in Fig.~\ref{Fig1}.
The system is described by the following Hamiltonian:
\begin{align}
&H=\omega_0 J_z+\omega_{\rm{a}} a^\dagger a+H_0\left(h\right)+H_{\rm I},\label{Total-H}\\
&H_{0}\left(h\right)=-\frac{\lambda}{2}\sum_{i=1}^{N}{\left[\frac{1\!+\!\gamma}{2}\sigma^x_i\sigma^x_{i+1}\!+\!\frac{1\!-\!\gamma}{2}\sigma^y_i\sigma^y_{i+1}\right]}\!-\!\frac{h}{2}\sum_{i=1}^{N}\sigma^z_i,\nonumber\\
&H_{\rm I}=g\left(a^\dagger J_-+aJ_+\right),\nonumber
\end{align}
%
where $H_{\rm I}$ represents the interaction Hamiltonian governing the atom-light coupling and the $XY$ Hamiltonian $H_0\left(h\right)$ characterizes interactions between nearest-neighbor spins.
Here, $J_{x,y,z}\!=\!\sum_{j=1}^{N}{\sigma_j^{x,y,z}/2}$ and $J_\pm\!=\!J_x\pm \rmi  J_y$ are the collective spin operators.
The operators $a^\dagger$ and $a$ correspond to the creation and annihilation of cavity mode photons.
The parameters $\omega_0$, $\omega_{\rm{a}}$, and $2g$ denote the spin transition frequency, cavity frequency, and single-photon Rabi frequency, respectively.
$\lambda$ describes the strength of the nearest-neighbor interaction.
$\gamma$ quantifies the degree of anisotropy.
$h$ represents the magnetic field to be estimated.
The experimental implementation of this generalized TC model is feasible in a superconducting quantum processor~\cite{PhysRevLett.120.050507,doi:10.1126/science.aay0600}.

However, it is important to note that the Hamiltonian~(\ref{Total-H}) cannot be solved exactly. 
To overcome this challenge, we introduce the time-averaged method to derive an effective description of the original system.
Moreover, this approach will provide valuable insights into the relationship between metrological scaling and quantum phases.
Employing the unitary transformation $U\!=\!\exp\left\{-\rmi \left[\omega_0 J_z+\omega_{\rm{a}} a^\dagger a+H_0\left(h\right)\right]t\right\}$  and the large detuning condition  ($\left|\omega_0-h-\omega_{\rm{a}}\right|\gg\lambda$), the total Hamiltonian (\ref{Total-H}) in the interaction picture could be read as 
$H_{\rm int-pic}= g\left[J_-a^\dagger\mathrm{e}^{-\mathrm{i}\left(\Delta+\delta\right) t}+\rm{H.c.}\right]$ with a large detuning $\Delta\!\equiv\!\omega_0-h-\omega_{\rm{a}}$ and a small residue $\delta$.
Utilizing the time-averaged method given in Ref.~\cite{James2007}, the total Hamiltonian can be approximated as (see more details in Appendix):
\begin{equation}\label{H2}
H_{\rm eff}^{\rm (s)}\simeq H_0\left(h-\omega_0\right)+\frac{2g^2}{\Delta}J_za^\dagger a+\frac{g^2}{\Delta}J_+J_-,
\end{equation}
which is written in the Schr\"{o}dinger picture.
The essence of the time-averaged method is to eliminate high-frequency contribution and thus it can be viewed as a natural generalization of the rotating-wave approximation~\cite{TAM}.
This approximation holds under the conditions of $\Delta^2\gg g^2N$, $\Delta^2\gg g^2N^2/\bar n$, and $|\Delta|\gg \lambda$.
Here, $\bar n\equiv \langle a^\dag a\rangle$ is the average photon number and the atom-light coupling $g$ is weak.

Generally, the average photon number is much larger than the number of spins, i.e., $\bar n\gg\langle J_z\rangle\approx N$, then we can further reduce the Hamiltonian~(\ref{H2}) to obtain the following effective Hamiltonian: 
\begin{equation}\label{H-eff}
H_{\rm {eff}}=H_0\left(h-\omega_0\right)+\frac{2g^2}{\Delta}J_za^\dagger a,
\end{equation}
which is similar to the Hepp-Coleman model~\cite{Sun06}.
Note that the estimated parameter $h$ not only exists in the term $H_0(h-\omega_0)$ but also appears in the detuning parameter $\Delta=\omega_0-h-\omega_{\rm{a}}$.
In summary, the conditions ensuring the validity of this effective Hamiltonian~(\ref{H-eff}) are $\Delta^2\gg g^2N$,  $|\Delta|\gg \lambda$, and $\bar n\gg N$.

\begin{figure}[t]
	\centering
	\begin{minipage}{1\linewidth}
		\centering
		\begin{overpic}[width=0.9\linewidth]{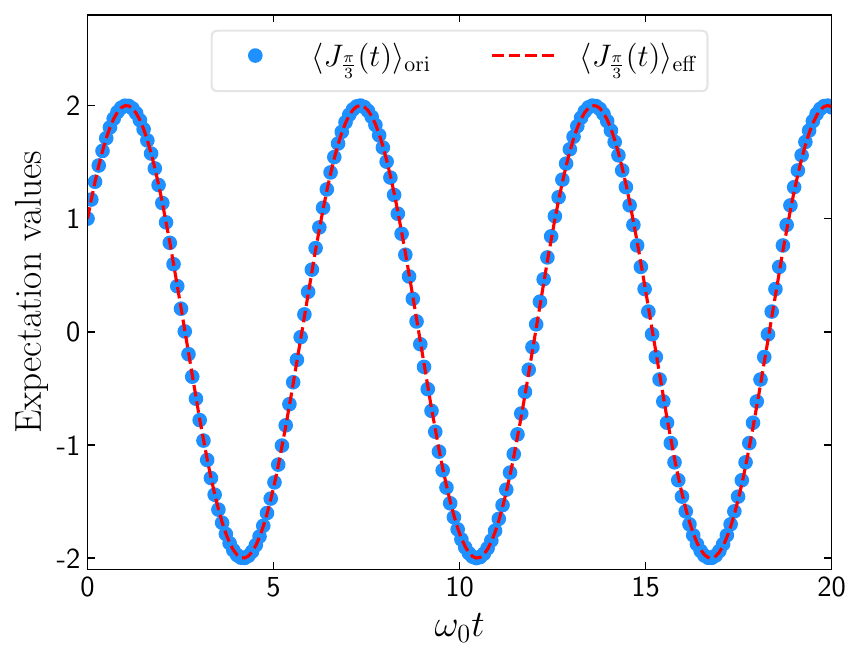}
		\end{overpic}
	\end{minipage}
	\caption{
(Color online)  
Comparison of expectation values derived from the original Hamiltonian~(\ref{Total-H}), indicated by the blue circles, and those obtained from the effective Hamiltonian~(\ref{H-eff}), represented by the red dashed line.
We  have employed the experimental parameters~\cite{WOS:000243867300038}: $\omega_0/\left(2\pi\right)\!=\!6.9$ GHz, $\omega_{\rm{a}}/\left(2\pi\right)\!=\!6.89$ GHz, $g/\left(2\pi\right)\!=\!1.05$ MHz.
Other parameters are $N\!=\!4$, $\bar{n}\!=\!40$, $\lambda\!=\!\gamma\!=\!1$, $h\!=\!10^{-5}$ Hz, $\theta\!=\!\pi/2$, $\phi=0$ and $\varphi\!=\!\pi/3$. 
The numerical computations have been conducted using the QuTip software package~\cite{JOHANSSON20121760}.
	}\label{Fig2}
\end{figure}

To demonstrate the validity of our effective Hamiltonian approximation, we perform calculations for the dynamics of the expectation value  $\left\langle M\left(t\right)\right\rangle $, where the observable $M$ is chosen as $J_{\varphi}=J_{x}\cos\varphi+J_{y}\sin\varphi$.
The initial state is considered as a product state, with the cavity initialized in a coherent state $\left|\alpha\right\rangle $ and the spins in a spin-coherent state $\ket{\theta,\phi}=\otimes_{i=1}^{N}(\cos\frac{\theta}{2}\ket{\uparrow}_i+e^{{\rm i}\phi}\sin\frac{\theta}{2}\ket{\downarrow}_i)$, where $\ket{\uparrow}_i$ and $\ket{\downarrow}_i$ are the eigenstates of $\sigma_i^z$ with the eigenvalues $1$ or $-1$, respectively. 
The consistency of the numerical expectations of the original and effective Hamiltonians (blue circles and red dashed line), as shown in Fig.~\ref{Fig2}, demonstrates that the effective Hamiltonian can faithfully describe the original system.
%
%

\section{Quantum Fisher information of the generalized TC model}\label{Sec.III}
The metrological scheme constitutes the initial preparation of the spins in the ground state $|\phi_{\rm g}\rangle$ of the $XY$ Hamiltonian $H_0(h)$, followed by the injection of a coherent light $\ket{\alpha}$. 
As a consequence, the initial state takes the form of a product state, denoted as $\ket{\psi(0)}=|\phi_{\rm g}\rangle\ket{\alpha}$.
Following quantum dynamics, the information about the magnetic field $h$ becomes encoded in the evolved state $\ket{\psi(t)}=\exp(-{\rm i}Ht)\ket{\psi(0)}$.
The maximum precision for estimating the parameter $h$ that can be provided by $\ket{\psi(t)}$ is determined by the QCRB: $\delta h\geq 1/\sqrt{\mathcal{F}_{h}}$, where $\mathcal{F}_{h}$ is the QFI~\cite{PhysRevLett.72.3439,helstrom1969quantum,holevo2011probabilistic,PhysRevD.23.357}:
\begin{equation}
\mathcal{F}_{h}=4[\ex{\psi(0)|\mathcal H^2|\psi(0)}-\ex{\psi(0)|\mathcal H|\psi(0)}^2].\label{QFI}
\end{equation}
Here, the metrological generator is given by
$
\mathcal{H}=\rmi \left(\partial_{h}U^{\dagger}\right)U
$ with $U=\exp(-{\rm i}Ht)$.
Having verified the validity of the effective Hamiltonian, we can replace $H$ with $H_{\rm eff}$~(\ref{H-eff}) to derive the generator as follows:
\begin{eqnarray}\label{generator}
\mathcal H=\left(1-\frac{2g^{2}}{\Delta^{2}}a^{\dagger}a\right)J_{z}t.
\end{eqnarray}
Substituting the generator~(\ref{generator}) into Eq.~(\ref{QFI}) we obtain
\begin{eqnarray}\label{QFI-2}
\mathcal F_h
& = & 4t^{2}\left[ \left(1-\frac{2g^{2}}{\Delta^{2}}\bar{n}\right)^{2}{\rm Var}(J_z)
+\frac{4g^{4}}{\Delta^{4}}\bar{n} \ex{J_{z}^{2}} \right],
\end{eqnarray}
where $\bar{n}\equiv\left|\alpha\right|^{2}$ is the average photon number of the coherent light $\ket{\alpha}$, $\ex{\cdot}$ denotes the expectation value over the ground state $|\phi_{\rm g}\rangle$ of the $XY$ model, and ${\rm Var}(J_z)=\ex{J_z^2}-\ex{J_z}^2$ is the variance.

Recall that we are exclusively considering the scenario where the average photon number $\bar n$ significantly exceeds the number of spins $N$.
As a result, in this context, the HL-precision refers to $\mathcal F_h\propto\bar n^2$.
Equation~(\ref{QFI-2}) makes it evident that the presence of nonvanishing spin fluctuations ${\rm Var}(J_z)$ is both necessary and sufficient to achieve the HL-precision in our metrological scheme.
Next, we will calculate the spin correlation functions, specifically ${\rm Var}(J_z)$ and $\ex{J_z^2}$, for the subsequent discussion.

   \begin{figure}[tpb]
	\centering
	\begin{minipage}{1\linewidth}
		\centering
		\begin{overpic}[width=0.7\linewidth]{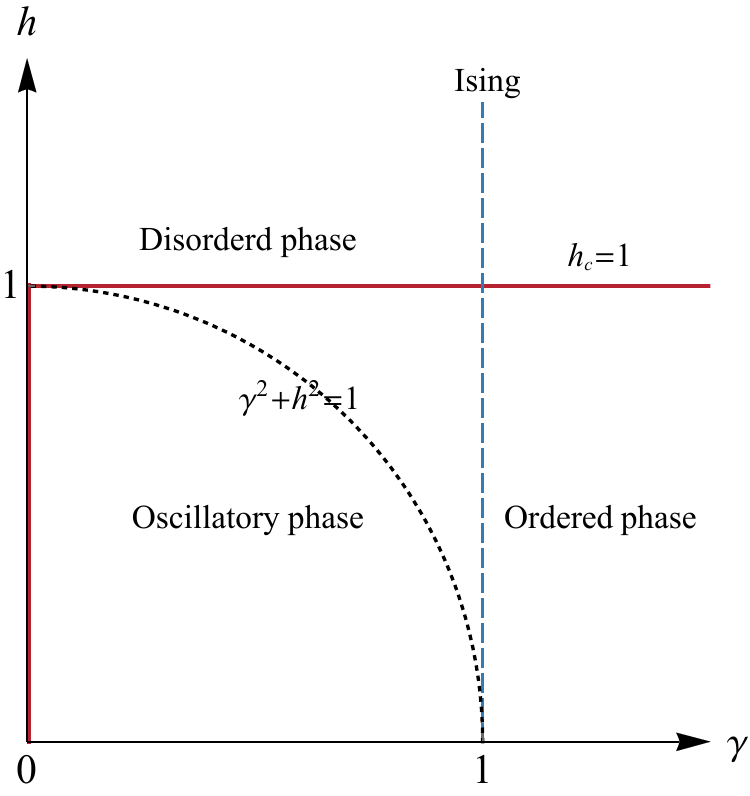}
		\end{overpic}
	\end{minipage}
	\caption{
		(Color online) 
		Phase diagram of the $XY$ model for $\gamma\geq0$ and $h\geq0$ under the normalization condition $\lambda=1$.
		The model exhibits critical behavior along the red line. 
		The critical magnetic field $h_{\rm c}\!=\!1$ separates the paramagnetic and the ferromagnetic phases. 
		The blue dashed line corresponds to the Ising model.
		On the black dots $\gamma^2+h^2\!=\!1$, the ground state can be factorized as a product state.
	}\label{Fig3}
\end{figure}

\subsection{Correlation functions in the $XY$ model}\label{Sec3-1}
To calculate the correlation functions, we can express the $XY$ model as a free fermion model by using the standard Jordan-Wigner transformation, Fourier transformation, and Bogoliubov transformation~\cite{LIEB1961407}.
Finally, the Hamiltonian of the spin part can be rewritten as 
\begin{equation}
H_{0}\left(h\right)=\sum_{k}\Lambda_{k}b_{k}^{\dagger}b_{k}+{\rm const},\label{DiaH}
\end{equation}
where the momentum is denoted as $k=2\pi m/N$ with $m=-N/2+1,\dots,N/2$ and $b_{k}$ represents the fermion  operator.
The excitation energy is given by
\begin{equation}
\Lambda_{k}=\sqrt{\left(h-\lambda\cos k\right)^{2}+\lambda^{2}\gamma^{2}\sin^{2}k},
\end{equation}
where the Bogoliubov angles are determined by
\begin{eqnarray}\label{angle}
\sin\nu_{k}=\frac{\lambda\gamma\sin k}{\Lambda_{k}},\quad
\cos\nu_{k}=\frac{\lambda\cos k-h}{\Lambda_{k}}.
\end{eqnarray} 
In terms of the $b_k$ operator, the total spin operator can be expressed as 
\begin{align}\label{Jz}
J_z&=\!-\frac{N}{2}+\frac{1}{2}\!\sum_{k}\!\left[\!(1+\cos\nu_k)b_k^\dag b_k\!\right.\nonumber\\
&\quad\left.+(1\!-\!\cos\nu_k)b_{-k}b_{-k}^\dag\!+\!{\rm i}\sin\nu_k (b_k^\dag b_{-k}^\dag\!-\! b_{-k}b_k)\!\right].
\end{align}
%
Based on Eq.~(\ref{Jz}) and the relation $b_k|\phi_{\rm g}\rangle=0$, we have 
\begin{align}
\ex{J_z^2}&=\frac{1}{2}\sum_k\sin^2\nu_k+\frac{1}{4}\left(\sum_k\cos\nu_k\right)^2,\label{Cor-Func1}\\
{\rm Var}(J_z)&=\frac{1}{2}\sum_k\sin^2\nu_k.\label{Cor-Func2}
\end{align}
%

The phase diagram of the $XY$ model is given in Fig.~\ref{Fig3}, where the paramagnetic and the ferromagnetic phases are separated by the critical magnetic field $h_{\rm c}=\lambda=1$~\cite{CAROLLO20201}.
Our objective is to elucidate the impact of the $XY$ interaction on the scaling behavior of the QFI based on Eqs.~(\ref{QFI-2}), (\ref{angle}), (\ref{Cor-Func1}), and (\ref{Cor-Func2}).

 \begin{figure}[tpb]
	\centering
	\begin{minipage}{1\linewidth}
		\centering
		\begin{overpic}[width=0.9\linewidth]{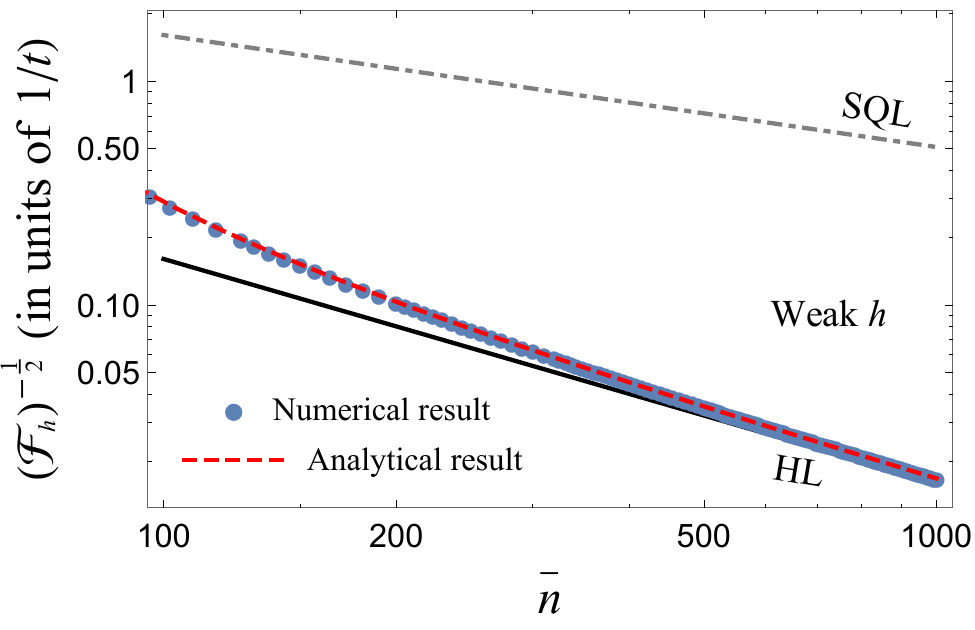}
			\put(0,63){(a)}
		\end{overpic}\\
		\begin{overpic}[width=0.9\linewidth]{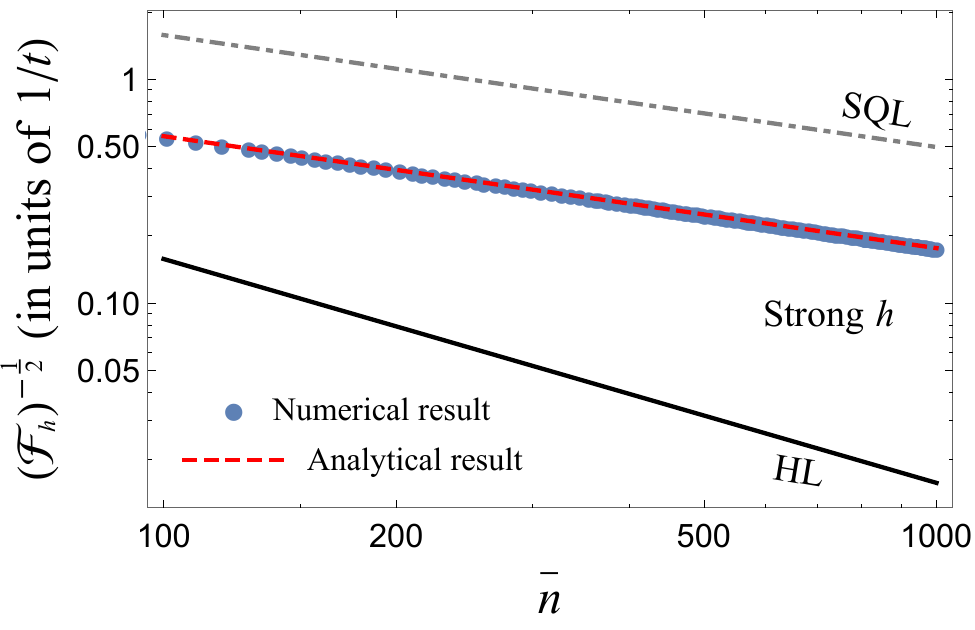}
			\put(0,63){(b)}
		\end{overpic}
	\end{minipage}
	\caption{
		(Color online) 
		Comparison of numerical and analytic results of sensitivity $1/\sqrt{\mathcal{F}_h}$ (in units of $1/t$) for the TC model with the Ising interaction:
		(a) the weak field case $h=10^{-5}$ Hz and (b) the strong field case $h=10^{5}$ Hz.
		The blue circles represent the numerical result obtained from Eq.~(\ref{QFI-2}), while the red dashed line denotes the analytic sensitivity given by Eqs.~(\ref{Ising-1}) and (\ref{Ising-2}).
		The gray dot-dash line indicates the SQL-precision ($\propto \bar n$) while the black line represents the HL-precision ($\propto \bar{n}^2$).
		In both cases, $N\!=\!8$ and the other parameters are the same as those in Fig.~\ref{Fig2}.
	}\label{Fig4}
\end{figure}

\subsection{Quantum Fisher information for the TC model}\label{Sec3-2}

We first consider the case of $\lambda\!=\!0$.
In this case, the Hamiltonian~(\ref{Total-H}) reduces to the TC model where the spins do not have direct interactions.
Substituting the condition $\lambda\!=\!0$ into Eqs.~(\ref{angle}),  (\ref{Cor-Func1}), and (\ref{Cor-Func2}), we obtain $\ex{J_z^2}=N^2/4$ and ${\rm Var}(J_z)=0$.
This is because the spin ground state is now fully polarized, $|\phi_{\rm g}\rangle\!=\!\ket{\uparrow,\uparrow,\dots,\uparrow}$.
Consequently,  from Eq.~(\ref{QFI-2}), we find that the QFI for the TC model is given by
\begin{eqnarray}
 \mathcal{F}_{h}(\lambda\!=\!0)\approx\frac{4g^4}{\Delta^4}t^{2} N^2 \bar n,
\end{eqnarray}
which indicates that only the SQL-precision, i.e., $\mathcal F_h\propto\bar n$, can be achieved.
Hence, our attention will shift to the impact of the spin interaction.

\subsection{Quantum Fisher information for the TC model with the Ising interaction}\label{Sec.3-3}

In this subsection, we focus on the TC model with the Ising interaction, i.e., $\gamma=1$.
For convenience, we set $\lambda=1$ in what follows.
We know that the Ising model will undergo a quantum phase transition from the ferromagnetic phase to the paramagnetic phase phase.
The corresponding quantum critical point is located at $h_{\rm c}=1$.
Our primary concern is how the QFI behaves in different quantum phases.
 
In the case of $h\ll h_{\rm c}$, we can approximate the Bogoliubov angles~(\ref{angle}) as $\sin\nu_{k}\approx\sin k$ and $\cos\nu_{k}\approx\cos k$.
Substituting this approximation into Eqs.~(\ref{Cor-Func1}) and (\ref{Cor-Func2}), we obtain 
 \begin{eqnarray}
&&\ex{J_z^2}\!\approx\!\frac{N}{4\pi}\!\int_{-\!\pi}^\pi\! dk\sin^2\!k\!+\!\frac{1}{4}\!\left(\!\frac{N}{2\pi}\!\int_{-\!\pi}^\pi\! dk\cos k\!\right)^{\!2}\!=\!\frac{N}{4},\\ 
&&{\rm Var}(J_z)=\frac{1}{2}\left(\frac{N}{2\pi}\int_{-\pi}^\pi dk\sin^2k\right)=\frac{N}{4},
\end{eqnarray}
where the thermodynamic limit has been considered. 
Compared to the scenario without spin interactions ($\lambda=0$), the Ising interaction induces a non-zero spin fluctuation, ${\rm Var}(J_z)=N/4$, for the weak magnetic field case.
As a result, the QFI is given by
\begin{eqnarray}\label{Ising-1}
 \mathcal{F}_{h}(\gamma\!=\!1,h\!\ll\!h_{\rm c})
 &\approx&
 4t^{2}\left[\left(1-\frac{2g^{2}}{\Delta^{2}}\bar{n}\right)^{2}\frac{N}{4}+\frac{4g^{4}}{\Delta^{4}}\bar{n}\left(\frac{N}{4}\right)\right]\n 
 &\approx&
 \frac{4g^{4}}{\Delta^{4}}t^{2}N\bar{n}^{2}
 +\mathcal O(\bar n),
\end{eqnarray}
which implies that the HL-precision for the photon number $\bar n$ can be achieved.

  \begin{figure}[tpb]
 	\centering
 	\begin{minipage}{1\linewidth}
 		\centering
 		\begin{overpic}[width=0.9\linewidth]{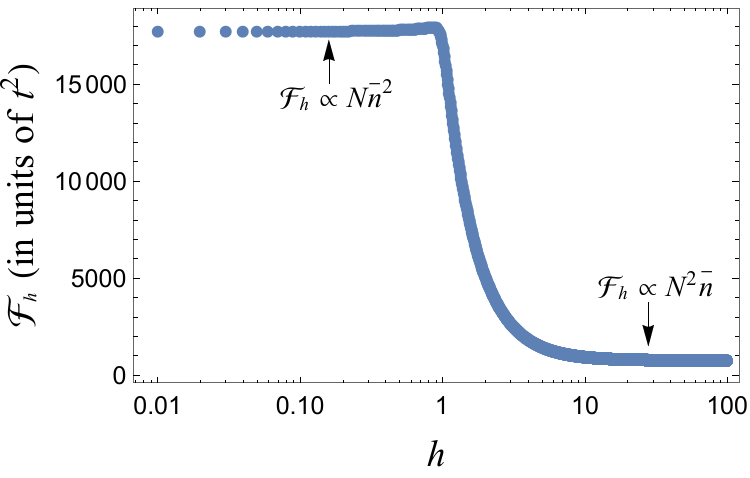}
 		\end{overpic}
 	\end{minipage}
 	\caption{
 (Color online) 
QFI (in units of $t^2$) versus the magnetic field $h$ in the TC model with the Ising interaction.
The blue circles are the numerical value of QFI obtained from Eq.~(\ref{QFI-2}).
Here, the number of spins is $N=40$ and the average photon number is $\bar{n}=1000$.
The other parameters are the same as those in Fig.~\ref{Fig2}.
 	}\label{Fig5}
 \end{figure}
 
However, in the case of $h\gg h_{\rm c}$, the approximation from Eq.~(\ref{angle}) $\sin \nu_k\approx0$ and $\cos\nu_k\approx -1$ results in $\ex{J_z^2}\approx N^2/4$ and ${\rm Var}(J_z)\approx 0$ which is consistent with the $\lambda=0$ case, see Sec.~\ref{Sec3-2}.
It should be emphasized that this approximation only requires $h\gg h_{\rm c}$ and also holds for the case when $\gamma\neq 1$.
Therefore, we can deduce from Eq.~(\ref{QFI-2}) that for $h\gg h_{\rm c}$ case, the QFI is the same as the one obtained from the TC model without any interaction, i.e., 
\begin{eqnarray}\label{Ising-2}
 \mathcal{F}_{h}(h\!\gg\!h_{\rm c})\approx\frac{4g^4}{\Delta^4}t^{2} N^2\bar n.
\end{eqnarray}

The analytical scalings of the QFIs (\ref{Ising-1}) and (\ref{Ising-2}) are numerically verified in Fig.~\ref{Fig4}, suggesting that by introducing spin interactions, our scheme can provide the HL-precision for estimating a weak magnetic field.
For the case of a strong magnetic field, the vanishing spin fluctuation ${\rm Var}(J_z)$ results in only achieving the SQL-precision.
Furthermore, figure~\ref{Fig5} illustrates a clear transition in the scaling of the QFI as we gradually increase the strength of the magnetic field $h$, ultimately crossing the quantum critical point $h_{\rm c}\!=\!1$.

 \subsection{Quantum Fisher information for the TC model with the $XY$ interaction}\label{Sec.3-4}
As discussed in Sec.~\ref{Sec.3-3}, we cannot surpass the SQL-precision for estimating a strong magnetic field in our metrological scheme. 
However, it is attainable for the weak field case by introducing the Ising interaction among the spins.
Therefore, in this subsection, we will mainly focus on the weak field case and illustrate the role of the spin anisotropy $\gamma$ within the $XY$ interaction played in the estimation precision.

\begin{figure}[tpb]
	\centering
	\begin{minipage}{1\linewidth}
		\centering
		\begin{overpic}[width=0.9\linewidth]{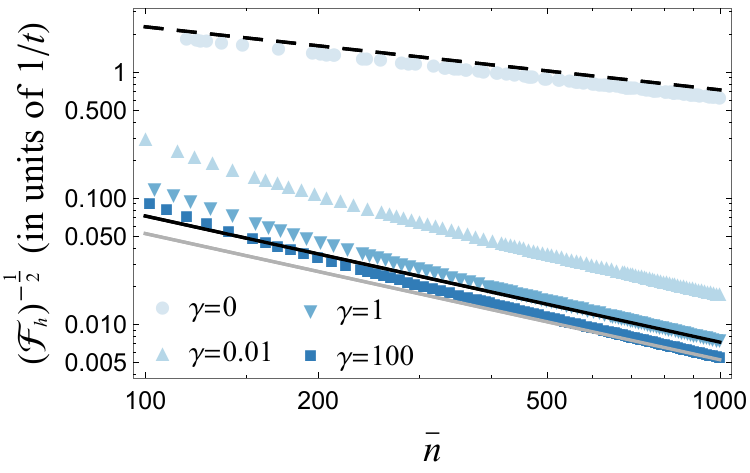}
			\put(0,63){(a)}
		\end{overpic}\\
		\begin{overpic}[width=0.9\linewidth]{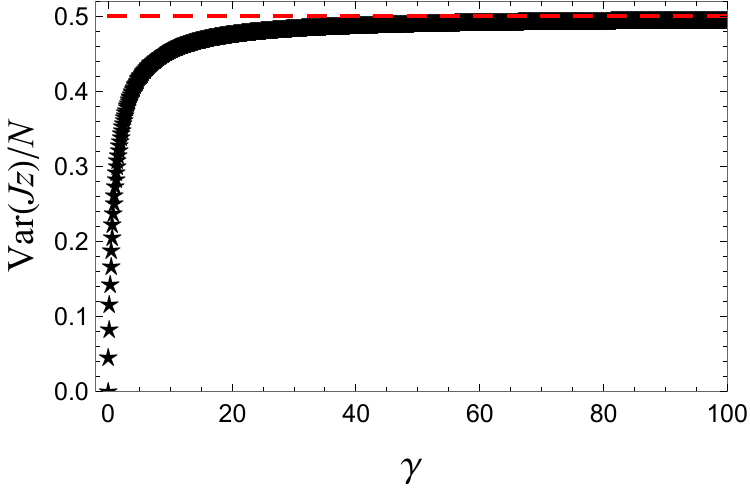}
			\put(0,63){(b)}
		\end{overpic}
	\end{minipage}
	\caption{(Color online) 
		(a) Comparison of numerical and analytic results of sensitivity $1/\sqrt{\mathcal{F}_h}$ (in units of $1/t$) for the TC model with the $XY$ interaction.
		The circles, upper triangles, lower triangles, and squares indicate the numerical results for $\gamma\!=\!0$, $\gamma\!=\!0.01$, $\gamma\!=\!1$, and $\gamma\!=\!100$, respectively.
		The black dashed line, black solid line, and gray solid line denote the analytical results, Eqs.~(\ref{g-0-h<hc}), (\ref{Ising-1}), and (\ref{g>>1}), for $\gamma\!=\!0$, $\gamma\!=\!1$, and $\gamma\!\gg\!1$.
		(b) Numerical result of the rescaled variance ${\rm Var}(J_z)/N$ versus the anisotropy parameter $\gamma$.
		The number of spins is taken as $N=40$.
		The magnetic field is chosen as $h\!=\!0.1$ and other parameters are the same as those in Fig.~\ref{Fig2}.
	}\label{Fig6}
\end{figure}

For the isotropic $XX$ model ($\gamma\!=\!0$), the Bogoliubov angles~(\ref{angle}) are given by $\sin\nu_k=0$ and $\cos\nu_k={\rm sgn}(\cos k-h)$, where we also assume $\lambda\!=\!1$ for convenience.
By Eqs.~(\ref{QFI-2}), (\ref{Cor-Func1}) and (\ref{Cor-Func2}), we know the spin fluctuation vanishes ${\rm Var}(J_z)=0$  and can conclude that the HL-precision cannot be achieved in this case. 
%
Explicitly, if $h>h_{\rm c}=1$ then we have $\cos\nu_k=-1$, and  as a result, $\ex{J^2_z}=N^2/4$.
Thus, the QFI~(\ref{QFI-2}) is given by
\begin{eqnarray}
\mathcal{F}_{h}(\gamma\!=\!0,h\!>\!h_{\rm c})=\frac{4g^4}{\Delta^4}t^{2} N^2 \bar n.
\end{eqnarray}
If $h<h_{\rm c}$, then by substituting $\sin\nu_k=0$ and $\cos\nu_k={\rm sgn}(\cos k-h)$ into Eqs.~(\ref{Cor-Func1}) and (\ref{Cor-Func2}) we obtain
\begin{eqnarray}
\ex{J_z^2}&=&\frac{1}{2}\sum_k\sin^2\nu_k+\frac{1}{4}\left(\sum_k\cos\nu_k\right)^2\n 
&\stackrel{N\to\infty}{=}&\frac{N^2}{16\pi^2}\!\left(\!-\!\int_{-\!\pi}^{-\!\arccos(h)}\!\!\!\!\!\!\!dk\!+\!\int_{-\!\arccos(h)}^{\arccos(h)}\!\!\!\!\!\!\!dk\!-\!\int^{\pi}_{\arccos(h)}\!\!\!\!\!\!\!dk\!\right)^2\n 
&=&\frac{N^2}{4\pi^2}[2\arccos(h)-\pi]^2.
\end{eqnarray}
%
Finally, from Eq.~(\ref{QFI-2}), the QFI is given by
\begin{eqnarray}\label{g-0-h<hc}
\mathcal{F}_{h}(\gamma\!=\!0,h\!<\!h_{\rm c})=\frac{[2\arccos(h)-\pi]^2}{\pi^2}\frac{4g^4}{\Delta^4}t^{2} N^2 \bar n,
\end{eqnarray}
which is verified in Fig.~\ref{Fig6}(a). 
Both of these two quantum phases only support the SQL-precision when $\gamma\!=\!0$.
This QFI scaling is the same as the one obtained in the TC model without spin interactions ($\lambda=0$).

In the region $0<\gamma<1$, no evident approximation is feasible, prompting us to resort to numerical calculations.
As depicted in Fig.~\ref{Fig6}(a), the emergence of the HL-precision is apparent even with a small anisotropy, i.e., $\gamma=0.01$.
This phenomenon can be understood from Fig.~\ref{Fig6} and Eq.~(\ref{QFI-2}) that the presence of nonvanishing spin fluctuation, i.e., ${\rm Var}(J_z)\neq0$ for $\gamma\neq0$, ensures the manifestation of such HL-precision.

For the $\gamma=1$ case (the Ising model), we have discussed in Sec.~\ref{Sec.3-3} and the HL-precision can be achieved in the weak field case.
If we increase the anisotropy to infinity $\gamma\!\gg\!1$ then Eq.~(\ref{angle}) can be approximated as $\sin\nu_k\approx{\rm sgn}(\sin k)$ and $\cos\nu_k\approx 0$.
Then, by Eqs.~(\ref{Cor-Func1}) and (\ref{Cor-Func2}), we have $\ex{J_z^2}\approx N/2$ and ${\rm Var}(J_z)\approx N/2$.
Thus, the QFI is given by
\begin{eqnarray}\label{g>>1}
\mathcal{F}_{h} (\gamma\!\gg\!1)
& \approx &  4t^{2}\left[\left(1-\frac{2g^{2}}{\Delta^{2}}\bar{n}\right)^{2}\frac{N}{2}+\frac{4g^{4}}{\Delta^{4}}\bar{n}\frac{N}{2}\right]\n
&\approx&\frac{8g^4}{\Delta^4}t^{2} N \bar n^2+\mathcal O(\bar n),
\end{eqnarray}
which has been verified in Fig.~\ref{Fig6}(a).
Figure~\ref{Fig6}(b) demonstrates that the spin fluctuation ${\rm Var}(J_z)$ increases with the raise of the anisotropy $\gamma$ and ultimately converges to the limiting value $N/2$.
Consequently, the QFI exhibits a behavior akin to ${\rm Var}(J_z)$ due to Eq.~(\ref{QFI-2}) and attains its maximum at sufficiently large values of $\gamma$.

\section{Conclusion}\label{Sec.V}
In conclusion, our study systematically delves into the pivotal role played by the $XY$ spin interaction in magnetic field sensing, employing a generalized TC model as the basis. 
The estimated precision is quantified by the QFI.
The effective description of the model is derived using the time-averaged method and rigorously validated through numerical calculations.
Based on the effective Hamiltonian, we establish a closed relationship between the QFI and the spin fluctuation, highlighting the indispensable nature of the spin fluctuation for achieving the HL-precision in estimation. 
Notably, in comparison to the TC model without any spin interactions, the introduction of the anisotropic $XY$ spin interaction beats the SQL and attains the HL-precision in estimating a weak magnetic field.
Furthermore, our investigation reveals a direct correlation between the QFI and the spin anisotropy, underlining the significant role of the spin anisotropy in realizing high-precision quantum metrology. 
Additionally, we observe a scaling transition of the QFI in the TC model with the Ising interaction. 
These findings not only contribute to quantum metrology within cavity-QED systems but also provide valuable insights for exploring many-body effect enhanced quantum metrology.

The time-averaged method we employed only reveals the mysteries of our generalized TC model within the large detuning region.
Distinct features in small detuning or other regions remain to be investigated.
We emphasize the possibility of employing another approach~\cite{LeonforteValentiSpagnoloCarolloCiccarello+2021+4251+4259,PhysRevA.97.042109} in the future by identifying the atoms as impurities to address such regions.

\section*{Acknowledgments}
This work is supported by the National Natural Science Foundation of China Key Grants No.~12134015 and No.~92365202.
Y.G.S. is supported by the National Natural Science Foundation of China (Grant No.~12247158), the ``Wuhan Talent'' (Outstanding Young Talents), and the Postdoctoral Innovative Research Post in Hubei Province.
W.J.L. is supported by the National Natural Science Foundation of China (Grant No.~12205092) and the Hunan Provincial Natural Science Foundation of China (Grant No.~2023JJ40208).
H.-L. S. was supported by the European Commission through the H2020 QuantERA ERA-NET Cofund in Quantum Technologies project “MENTA” and the Hefei National Laboratory.

\begin{appendix}
\setcounter{section}{1}
\setcounter{equation}{0}
\section*{Appendix:~Effective Hamiltonian of the cavity-QED system}\label{Appendix}
Here, we employ the time-averaged method~\cite{TAM} to derive an effective description for the original system~(\ref{Total-H}).
This method can be regarded as a generalized rotating wave approximation, which eliminates various high-frequency contributions.
It states that if we can express our Hamiltonian in the following form:
\begin{eqnarray}\label{H-int-2}
H_{\rm int-pic}=\sum_i f_i\exp(-\mathrm{i} \Omega_i t)+{\rm H.c.},
\end{eqnarray}
then we can approximate it as 
\begin{eqnarray}\label{H-int-3}
H_{\rm int-pic}\approx \sum_{i,j}{\frac{1}{\bar{\Omega}_{ij}}\left[f_i^\dagger,f_j\right]\mathrm{e}^{\mathrm{i}\left(\Omega_i-\Omega_j\right)t}}.
\end{eqnarray}
Here, $\bar\Omega_{ij}$ is the harmonic average of frequencies $\Omega_i$ and $\Omega_j$, i.e., 
$1/\bar \Omega_{ij}=\left(1/\Omega_i+1/\Omega_j\right)/2$.

In our case, the total Hamiltonian (\ref{Total-H}) in the interaction picture is given by
\begin{eqnarray}
H_{\rm int-pic}\label{H-int-1}
&=&
\mathrm{i}\frac{d U_1^\dag}{dt} U_1 +U_1^\dag HU_1  \nonumber \\ 
&=&g\left[\mathrm{e}^{\mathrm{i}\omega_{\rm{a}}t}a^\dagger\mathrm{e}^{ \mathrm{i}H_0\left(h-\omega_0\right)t}J_-\mathrm{e}^{-\mathrm{i}H_0\left(h-\omega_0\right)t}+\rm{H.c.}\right]\nonumber \\
&=& g\left[J_-a^\dagger\mathrm{e}^{-\mathrm{i}\left(\Delta+\delta\right) t}+\rm{H.c.}\right],
\end{eqnarray}
where $U_1\!=\!\exp\left\{-\mathrm{i}\left[\omega_0J_z+\omega_{\rm{a}} a^\dag a+H_0(h)\right]t\right\}$ is the
unitary transformation operator, $\left.\Delta=\omega_0-h-\omega_{\rm{a}}\right.$ is a large effective detuning, and a small residue $\delta$ comes from the commutators $[H_0(0), J_-]$ and $[H_0(0), J_z]$.
Under the large detuning condition ($|\Delta|\gg \lambda$), the Hamiltonian (\ref{H-int-1}) takes the form of the Hamiltonian~(\ref{H-int-2}) with $f_i=g\sigma_i^-a^\dag/2$ and $\Omega_i=\Omega_j\simeq\Delta$.
Then, using Eq.~(\ref{H-int-3}), we derive the effective Hamiltonian in the interaction picture:
\begin{align}\label{H-int-eff-4}
H_{\rm int-pic}
&\approx
\sum_{i,j=1}^{N}\frac{g^2}{4\Delta}\left[\sigma_i^+ a,\sigma^-_j a^\dag\right]\n
&= \frac{g^2}{\Delta}(J_{+}J_{-}+2J_{z}a^\dag a).
\end{align}
%
Rewriting Eq.~(\ref{H-int-eff-4}) into the Schr\"{o}dinger picture, we obtain the effective Hamiltonian: 
\begin{eqnarray}
H_{\rm eff}^{\rm (s)}
&\approx&
\omega_0J_z+\omega_{\rm{a}} a^\dag a+H_0(h)
+\frac{2g^2}{\Delta}J_za^\dagger a+\frac{g^2}{\Delta}J_+J_-\nonumber \\
&\approx&
 H_0\left(h-\omega_0\right)+\frac{2g^2}{\Delta}J_za^\dagger a+\frac{g^2}{\Delta}J_+J_-,
\end{eqnarray}
where the term $a^\dag a$ has been omitted since the photon number remains conserved.

To ensure the validity of this approximation, in addition to the condition $|\Delta|\gg\lambda$, we also require that the timescale of the first term in $H_{\rm eff}$ be greater than the timescale introduced by the corrections (the second and the third terms)~\cite{PhysRevA.102.052615}.
To compare the magnitude of these terms, it is convenient to express the effective Hamiltonian in the rotating frame by employing the unitary transformation $U_{2}=\exp\left[{-\mathrm{i}\omega_0\left(J_z+a^\dagger a\right)t}\right]$, resulting in
\begin{align}
H_{\rm{eff}}^{\left(\rm{r}\right)}&=\mathrm{i}\frac{dU_{2}^{\dagger}}{dt}U_{2}+U_{2}^{\dagger}H_{\rm{eff}}^{\rm (s)}U_{2}\n
&\approx-\Delta a^{\dagger}a+\frac{2g^{2}}{\Delta}J_{z}a^{\dagger}a+\frac{g^{2}}{\Delta}J_{+}J_{-},
\end{align}  
which suggests that the first term scales as $|\Delta| \bar n$, the second term scales as $g^2N\bar n/|\Delta|$, and the third term scales at most $g^2N^2/|\Delta|$.
Thus, the conditions ensuring the validity of the approximation are $|\Delta|\bar n\gg g^2N\bar n/|\Delta|$, $|\Delta|\bar n\gg  g^2N^2/|\Delta|$, i.e., $\Delta^2\gg g^2N$, $\Delta^2\gg g^2N^2/\bar n$, and the large detuning condition ($|\Delta|\gg \lambda$).
Here, $\bar n=\langle a^\dag a\rangle$ is the average photon number. 

\end{appendix}



%

\end{document}